\documentclass[pra,aps,a4paper,twocolumn,showpacs,amsmath,amssymb,floatfix,nobalancelastpage,superscriptaddress,notitlepage,accepted=2021-09-03]{quantumarticle}
\pdfoutput=1
\usepackage{color}
\usepackage{bbm}
\usepackage{graphicx}
\usepackage{dcolumn}
\usepackage[utf8]{inputenc}

\usepackage[shortlabels]{enumitem}
\usepackage{graphicx}
\usepackage{epstopdf}
\usepackage{amsthm}
\usepackage{amsmath}
\usepackage{empheq}
\usepackage{bbm}
\usepackage{braket}
\usepackage{amssymb}
\usepackage{pifont}
\usepackage[usenames,dvipsnames]{xcolor}
\usepackage{cases}
\usepackage{latexsym}

\usepackage[breaklinks,colorlinks=true,citecolor=Cerulean,linkcolor=RubineRed,urlcolor=Cerulean]{hyperref}
\usepackage{subfigure}
\usepackage{verbatim}
\usepackage{soul}

\graphicspath{ {images/}{images/figure1/}{images/figure2/}{images/figure3/}{images/figure4/}{images/figure5/} }

\newtheorem{theorem}{Theorem}

\newtheorem{lemma}{Lemma}

\newcommand{\ketbra}[2]{|#1\rangle\langle #2|}

\newcommand{\id}{\mathbbm{1}} 
\usepackage{lipsum}
\usepackage{graphicx}

\newcommand{\beq}{\begin{equation}}
\newcommand{\enq}{\end{equation}}
\newcommand{\beqal}{\begin{equation}\begin{aligned}}
\newcommand{\enqal}{\end{aligned}\end{equation}}
\newcommand{\beqst}{\begin{equation*}}
\newcommand{\enqst}{\end{equation*}}
\newcommand{\beqar}{\begin{eqnarray}}
\newcommand{\enqar}{\end{eqnarray}}
\newcommand{\beqarst}{\begin{eqnarray*}}
\newcommand{\enqarst}{\end{eqnarray*}}
\newcommand{\beit}{\begin{itemize}}
\newcommand{\enit}{\end{itemize}}

\mathchardef\mhyphen="2D

\newcommand{\suppress}[1]{}

\newcommand{\red}[1]{{\color{red}#1}}

\newcommand{\mb}[1]{\mathbb{#1}}

\newcommand{\Tr}{\mathrm{Tr}} 
\newcommand{\supp}{\mathrm{supp}}

\newcommand{\diag}{\mathrm{diag}}

\newcommand{\Z}{\mb{Z}}

\newcommand{\ketn}[1]{| #1 \rangle}
\newcommand{\bran}[1]{\langle #1 |}
\newcommand{\braketn}[2]{\langle #1 | #2 \rangle}
\newcommand{\ketbran}[2]{\ketn{#1} \! \bran{#2}}
\newcommand{\sandwich}[3]
{\left\langle  #1 \right| #2 \left| #3 \right\rangle}

\begin{document}

\title{Computable R\'enyi mutual information: Area laws and correlations}

\author{Samuel O. Scalet}
\affiliation{Max-Planck-Institut für Quantenoptik, Hans-Kopfermann-Straße 1, D-85748 Garching, Germany}
\affiliation{Munich Center for Quantum Science and Technology (MCQST), Schellingstr. 4, D-80799 München, Germany}
\author{\'Alvaro M. Alhambra}
\affiliation{Max-Planck-Institut für Quantenoptik, Hans-Kopfermann-Straße 1, D-85748 Garching, Germany}
\affiliation{Munich Center for Quantum Science and Technology (MCQST), Schellingstr. 4, D-80799 München, Germany}
\author{Georgios Styliaris}
\affiliation{Max-Planck-Institut für Quantenoptik, Hans-Kopfermann-Straße 1, D-85748 Garching, Germany}
\affiliation{Munich Center for Quantum Science and Technology (MCQST), Schellingstr. 4, D-80799 München, Germany}
\author{J. Ignacio Cirac}
\affiliation{Max-Planck-Institut für Quantenoptik, Hans-Kopfermann-Straße 1, D-85748 Garching, Germany}
\affiliation{Munich Center for Quantum Science and Technology (MCQST), Schellingstr. 4, D-80799 München, Germany}
\begin{abstract}
The mutual information is a measure of classical and quantum correlations of great interest in quantum information. It is also relevant in quantum many-body physics, by virtue of satisfying an area law for thermal states and bounding all correlation functions. However, calculating it exactly or approximately is often challenging in practice. Here, we consider alternative definitions based on R\'enyi divergences. Their main advantage over their von Neumann counterpart is that they can be expressed as a variational problem whose cost function can be efficiently evaluated for families of states like matrix product operators while preserving all desirable properties of a measure of correlations. In particular, we show that they obey a thermal area law in great generality, and that they upper bound all correlation functions. We also investigate their behavior on certain tensor network states and on classical thermal distributions.
\end{abstract}

\maketitle

\section{Introduction}One of the most important features of quantum systems is the nature of their \textit{correlations}, which differ from their classical counterparts, and lie behind the complexity of many-body quantum states. It is known, however, that when these correlations are weak and spatially localized, one can devise efficient methods to classically simulate complex quantum states via, for instance, tensor network methods \cite{eisert2013entanglement,cirac2020}. This occurs at least in gapped ground states in 1D \cite{verstraete2006matrix,hastings2007area}, Gibbs states of local Hamiltonians \cite{hastings2006solving,molnar2015approximating,kuwahara2020}, and low-depth quantum circuits \cite{vidal2003efficient}. Characterizing and quantifying those correlations is hence a subject of wide interest in fields ranging from quantum computing to condensed matter physics. 

The \textit{quantum mutual information} is perhaps the most widely known measure of correlations in quantum systems. It has a number of desirable properties, such as positivity or being nonincreasing under local operations, and a well-defined information-theoretic meaning \cite{groisman2005}. For a bipartite state $\rho_{AB}$, it is given by
\beq \label{eq:mutS}
I(A:B) = S(\rho_A) +S(\rho_B) - S(\rho_{AB}),
\enq
where $S(\rho)=-\Tr[\rho\log\rho]$ is the von Neumann entropy.
Importantly, it is known to obey an area law for all quantum Gibbs states \cite{wolf2008,kuwahara2020}, which implies that the correlations between adjacent subsystems scale only like their mutual boundary and are thus spatially localized. {Calculating the mutual information in quantum systems is hence an important task in many physically relevant scenarios. However, this is often impossible via known analytical methods and requires numerically diagonalizing the whole density matrix, and no efficient methods to calculate it for matrix product operators are available.} 

It is thus highly desirable to find measures of correlations that share the appealing information-theoretic properties of $I(A:B)$, but are simpler to compute in practice, for instance, through variational algorithms. This is often done by replacing the entropies in Eq. \eqref{eq:mutS} with the more general $\alpha$-R\'enyi entropies. The resulting quantity can be computed via a variety of numerical and analytical means and has been shown to characterize phenomena such as quantum \cite{Alcaraz_2014,St_phan_2014} and thermal \cite{singh2011finite} phase transitions,  or the correlations in many-body localization \cite{banuls2017}. However, it lacks a number of important properties, which prevent it from being a sensible measure of correlations. In particular, it can be negative \cite{kormos2017} and can increase under local operations.

Motivated by this, we here explore alternative definitions of the R\'enyi mutual information, based on the notion of quantum R\'enyi divergences \cite{tomamichel2016,khatri2020}. These are measures of distinguishability of quantum states, which play a pivotal role in information-theoretic tasks, such as single-shot communication protocols \cite{berta2011,anshu2017}, channel coding \cite{wilde2014strong,Mosonyi_2015,Leditzky_2016,Mosonyi_2017,Ding_2018,fang2019geometric} or hypothesis testing \cite{Mosonyi_2014,hayashi2016}. In principle, each of the many variants of quantum R\'enyi divergences \cite{petz1986,muellerlennert2013,wilde2014strong,matsumoto2018,datta2009,audenaert2015,fawzi2021} allows us to define a mutual information as we explain in Appendix \ref{sec:divergences}. Here, we focus on two particular cases and explain how to compute them in practice, at least when the input state is represented via tensor networks. We show that they satisfy desirable properties of the mutual information, including an area law for thermal states, which constitutes our main technical result.  This area law holds (i) in one dimension, (ii) for high temperatures, (iii) for commuting Hamiltonians, or (iv) in classical states, where in the latter case it does not depend on the temperature. We also show that, like $I(A:B)$, they bound all correlation functions, and that one of them yields by construction an area law for a broad class of tensor network states. 
{Our results are summarized in Table \ref{table:results}.}
\begin{table}[t]
    \centering
    \begin{tabular}{|l|c|c|}
        \hline
         & $I_\infty(A:B)$ & $I_\alpha^\mathbb M(A:B)$ \\ \hline
        Nonnegative & \checkmark & \checkmark \\ \hline
        Nonincreasing under LO& \checkmark & \checkmark \\ \hline
        \underline{Thermal area laws:} & & \\
        One dimension & \checkmark & \checkmark \\ 
        High temperature & \checkmark & \checkmark \\
        Commuting & \checkmark & \checkmark \\
        Classical $\beta$-independent & \ding{55} & $\alpha=2$ \\ \hline
        PEPDO area law & \ding{55} & $\alpha=2$ \\ \hline
        Bound on correlations & \checkmark & \checkmark \\ \hline
    \end{tabular}
    \caption{Table of properties for the proposed R\'enyi mutual informations defined in Section \ref{sec:def}.}
    \label{table:results}
\end{table}

The manuscript is structured as follows. In Section \ref{sec:def}, we discuss the definitions of R\'enyi mutual information and their relevant variational expressions. In Section \ref{sec:areaLaw}, we show the thermal area law and explain the regimes in which it applies. Then, in Section \ref{sec:PEPO} we bound its behavior for certain tensor network states, and in Section \ref{sec:corr}, we explain that any one of these measures upper bounds all correlation functions. The technical proofs, as well as further details, can be found in the Appendix.

\section{Definitions of R\'enyi mutual information}\label{sec:def}We consider a quantum system on a finite lattice $\Lambda \subset \Z^D$ with local Hilbert spaces $\mathbb{C}^d$.
For a quantum state $\rho$, we denote its reduced density matrices on subsystems $A, B \subset \Lambda$ as $\rho_A$ and $\rho_B$ respectively.

Apart from \eqref{eq:mutS}, the quantum mutual information is also given by the following equivalent expressions
\begin{subequations}
\begin{align}
I(A:B) &= D(\rho_{AB}\| \rho_A\otimes \rho_B)\label{eq:mutD}\\
&=\min_{\sigma_B}D(\rho_{AB}\|\rho_A\otimes \sigma_B)\label{eq:mutOpt}
\end{align}
\end{subequations}
with the Umegaki relative entropy $D(\rho\|\sigma)=\Tr[\rho\log\rho-\rho\log\sigma]$.
This quantity is nonnegative and cannot increase under local operations on both $A$ and $B$. These properties follow, respectively, from the positivity of the relative entropy, and its contractivity under CPTP maps, i.e., the \textit{data-processing inequality}.
To obtain R\'enyi versions of the mutual information, one can then generalize any one of Eq. \eqref{eq:mutS}, \eqref{eq:mutD} and \eqref{eq:mutOpt}. However, each of them yields a different definition, which are no longer equivalent.

\subsection{R\'enyi entropies}
Starting from \eqref{eq:mutS} and replacing the von Neumann entropies in the definition of the mutual information by the R\'enyi entropies $S_\alpha(\rho)=(1-\alpha)^{-1}\log\Tr[\rho^\alpha]$, one obtains
\beq \label{eq:ialpha}
I_\alpha(A:B) = S_\alpha(\rho_A) +S_\alpha(\rho_B) - S_\alpha(\rho_{AB}).
\enq
For integer values of $\alpha$, this definition contains only integer powers and traces of the density matrices. This feature makes it easily computable in many physically relevant situations. Analytically, an important example is the \textit{replica trick} \cite{calabrese2004,calabrese2009}, which has been used to calculate $S_\alpha$ in a conformal field theory for integer values of $\alpha$ (see \cite{asplund2014mutual,agon2016quantum,Chen_2019} for calculations of $I_\alpha(A:B)$. {The same method has been used to calculate certain R\'enyi entropies and trace distances \cite{lashkari2014,zhang2019}}). It can also be calculated exactly for free fermions \cite{bernigau2015}. Numerically, it is efficiently computable when the state $\rho$ is represented by a matrix product density operator (MPDO) \cite{pirvu2010matrix}, or by quantum Monte Carlo methods \cite{cirac2010,hastings2010measuring,humeniuk2012quantum,grover2013entanglement}. 

However, this definition lacks several of the important properties of $I(A:B)$.
For instance, it can be negative in physically relevant situations \cite{kormos2017}. We show in appendix \ref{sec:ising} that this can, in fact, happen in a very simple scenario: the thermal state of a classical Ising chain with an external field. To do so, we calculate the mutual information arising from Eq.~\eqref{eq:ialpha} analytically in the thermodynamic limit using transition matrices and show that for a sufficiently weak antiferromagnetic coupling it takes negative values.
Moreover, its negativity implies that it cannot be nonincreasing under local operations in general: tracing out the $A$ system is a local operation that can increase $I_\alpha(A:B)$ from a negative value to zero.

\subsection{Maximal R\'enyi divergence}A possible strategy to obtain a R\'enyi mutual information, which inherits the desirable properties of $I(A:B)$, is to invoke one of its equivalent definitions in terms of relative entropy (also known as \textit{divergence}) and directly extend the latter to the R\'enyi case.
For applications in quantum communication, it is common to generalize Eq. \eqref{eq:mutOpt}, e.g., see \cite{khatri2020}. However, let us here start from Eq. \eqref{eq:mutD}. To do so, we introduce the important \textit{maximal} R\'enyi divergence \cite{renner2008security}.
Given quantum states $\rho$ and $\sigma$, it is defined as \cite{datta2009}
\beq
\label{eq:defDmax}
D_\infty(\rho\|\sigma)=\log\inf\{\lambda: \rho\le\lambda\sigma\},
\enq
which, in turn, yields a definition of a R\'enyi mutual information as $I_\infty(A:B)=D_\infty(\rho_{AB}\|\rho_A\otimes\rho_B)$. 
The latter is nonnegative and cannot increase under local operations, where the last property follows from the data-processing inequality of the divergence under CPTP maps $\mathcal E$~\cite{khatri2020}, i.e., $D_\infty(\mathcal E(\rho)\|\mathcal E(\sigma))\le D_\infty(\rho\|\sigma)$.

This quantity has two important features. The first is that it can be approximated when the arguments are matrix product operators. We can rewrite Eq. \eqref{eq:defDmax} as
\begin{align} \label{eq:max1}
D_\infty(\rho\|\sigma)&=\log\inf\{\lambda: \inf_{\ketn{\psi}} \sandwich{\psi}{\lambda\sigma-\rho}{\psi}\ge0\}.
\end{align}
It is then possible to approximate the braket in Eq. \eqref{eq:max1} using the DMRG algorithm \cite{white1992}. 
While convergence is not guaranteed, it can be checked whether it approaches a limit with increasing bond dimension of $\ketn\psi$, indicating that the {infimum} has been well approximated.
The minimal $\lambda$ can then be determined using a binary search.
A potential difficulty is that determining whether an MPO is positive is an NP-hard problem \cite{kliesch2014} and calculating $D_\infty$ involves determining the positivity of the MPO $\lambda\rho_A \otimes \rho_B -\rho$. As this problem has additional structure it might still be efficiently solvable in practice. 
{This is supported by its similarty with the usual target of the the DMRG algorithm, which is finding the lowest eigenvalue of a particular MPO $H$
\begin{align}
    \inf_{\ket{\psi}} \sandwich\psi H \psi.
\end{align}
This has been employed succesfully in many cases, despite also solving an instance of the NP-hard problem of positivity, as a nonnegative smallest eigenvalue is equivalent to a nonnegative state \cite{schollwock2005,schollwock2011,landau2015,block2021}. In fact, the ground state problem is even QMA-hard \cite{aharonov2009}.}

The second important feature is that it  upper bounds all R\'enyi divergences that fulfill the data-processing inequality \cite{tomamichel2016}. This is relevant since, as we show below, it follows a thermal area law, which automatically extends to all known divergences.

We briefly comment on the generalization using Eq. \eqref{eq:mutOpt} instead. This yields another possible definition of R\'enyi mutual information, which in fact has an operational interpretation in communication theory as the communication cost of entanglement-assisted one-shot communication protocols \cite{berta2011,anshu2017}.
However, the additional optimization makes the quantity more difficult to handle both computationally and analytically, and hence less suited for our present purpose.
Nevertheless, our area laws also apply to this quantity as it is by definition smaller than $I_\infty$.

\subsection{Measured R\'enyi divergence}
We now present another quantity that can be efficiently computed numerically via variational algorithms.
For classical states, i.e., probability distributions $P$ and $Q$, the R\'enyi divergence is defined as \cite{renyi1961} 
\beq
\label{eq:renyiDivClassical}
D_\alpha(P\|Q)=\frac{1}{\alpha-1}\log\sum_x P(x)^\alpha Q(x)^{1-\alpha}.
\enq
From this, the \textit{measured R\'enyi divergence} is defined as \cite{berta2017}
\beq
D^\mathbb{M}_\alpha(\rho\|\sigma)=\sup_{(\chi, M)} D_\alpha(P_{\rho, M}\|P_{\sigma, M}),
\enq
with a supremum over all possible POVM measurements \red{$M$} and $P_{., M}$ the post measurement states, i.e., the respective probability distributions over the measurement outcomes.
{$\chi$ is the set of measurement outcomes, whose size can vary.}
We denote the corresponding mutual information by $I_\alpha^\mathbb{M}(A:B)=D^\mathbb{M}_\alpha(\rho_{AB}\|\rho_A\otimes\rho_B)$.
Like $I(A:B)$ and $I_\infty(A:B)$, this mutual information is positive and does not increase under local operations.
Moreover, $D^\mathbb{M}_\alpha\le D_\infty$ for all $\alpha$, so all thermal area laws for $I_\infty(A:B)$ also apply to $I_\alpha^\mathbb{M}(A:B)$.

The interest of this quantity stems from the following variational expression for $\alpha>1$ \cite{berta2017}
\beq \label{eq:varMeasured}
D_\alpha^\mathbb{M}(\rho\|\sigma)=\frac{1}{\alpha-1}\log\sup_{\omega>0}\alpha\Tr[\rho\omega^{\alpha-1}]+(1-\alpha)\Tr[\sigma\omega^\alpha].
\enq
 Again assuming the states to be given as MPDOs, this is an optimization in which the target function only contains products and traces of MPDOs, which can be efficiently computed for integer values of $\alpha>1$ with DMRG-type algorithms. In general, this can be done by optimizing over the matrix product operators with purifications to enforce the positivity constraint. 

For $\alpha=2$, we give an explicit expression for the optimizer.
The positivity constraint can be dropped as there cannot be an optimizer with negative eigenvalues. Using a vectorized notation $\ketbran ij\to\ket i\ket j$, we then obtain (see Appendix \ref{sec:optimizer})
\beq
\label{eq:optimizer}
\ket{\omega}=\frac{1}{2} \frac{1}{\sigma\otimes\id+\id\otimes\sigma}\ket{\rho}.
\enq

A direct calculation of $\omega$ from this is inefficient due to the inverse involved, but one can instead determine $\omega$ variationally by minimizing $\| 2(\sigma\otimes\id+\id\otimes\sigma)\ket\omega-\ket\rho\|_2^2$, which is a quadratic expression in $\ket\omega$ and thus, again, it can be obtained using DMRG-like algorithms.
{For other values of $\alpha$ the derivative of Eq. \eqref{eq:varMeasured} is not linear in $\omega$ and hence no such simple expression for the optimizer can be given.}

{As we will see in the following sections, the 2-measured R\'enyi mutual informations also fulfills all of the properties in the following chapters and may therefore be considered to be the most useful out of the measured R\'enyi mutual informations.}

\section{Thermal area laws}\label{sec:areaLaw}
In this section we present our main technical results: area laws for the R\'enyi mutual information in thermal states of local Hamiltonians.
In the following, we consider a subset of a lattice A and its complement B.
The Hamiltonian is split into three parts
\begin{equation}
    H=H_A+H_B+H_I,
\end{equation}
such that $H_A$, $H_B$ have support in $A$ and $B$ respectively, and $H_I$ is the interaction term.
We consider thermal states $\rho=\exp(-\beta H)/Z$, where the partition function is defined as $Z=\Tr[e^{-\beta H}]$.
We assume that $H$ consists of local, finite-range terms $H=\sum h_i$, where each $h_i$ is supported on at most $k$ sites and is bounded such that $\sup_x\sum_{\supp(h_i)\ni x}\|h_i\|= J<\infty$.
Consequently, $\|H_I\|$, which contains only those local terms that have support in $A$ and $B$, scales with the size of the boundary.
We further define $\partial A=\{x\in A : \exists h_i\text{ s.t. }x\in\supp(h_i)\land B\cap \supp(h_i)\ne\emptyset\}$ as the boundary of $A$, which is such that there are no terms with support in both $A\setminus\partial A$ and $B$. Notice that in 1D, $\vert \partial A \vert \propto \text{const.}$, independent of the system size.


We first introduce a technical lemma, which allows us to prove area laws in several special cases.
\begin{lemma} \label{lem:geometric}Let $E_\beta=e^{-\beta (H_A+H_B)}e^{\beta H}$. For any Hamiltonian defined as above, the maximal R\'enyi mutual information of a thermal state fulfills
\beq
I_\infty(A:B)\le\beta \|H_I\|+\log\left(\|E_{\beta/2}\|^2 \|E_{\beta/2}^{-1}\|^4\right).
\enq
\end{lemma}
The proof is in Appendix \ref{sec:geometric}. While the first term of the RHS directly scales with the boundary and corresponds to the result in \cite{wolf2008}, one still needs to bound the additional second term.

In the case of a commuting Hamiltonian, we have $E_\beta = e^{\beta H_I}$ straightforwardly, and thus we find:
\begin{theorem}[Commuting Hamiltonian]\label{thm:commuting}
For $[H,H_A+H_B]=0$ it holds that
\beq
I_\infty(A:B)\le 4\beta  \|H_I\|.
\enq
\end{theorem}

For Hamiltonians with non-commuting terms, it is no longer possible to cancel the bulk contributions in $E_\beta$ directly, but the norm can still be bounded in many cases.

In 1D systems, we use a lemma from \cite{perezgarcia2020} (section 2.3.1 and Theorem 3.1) based on previous work by Araki \cite{araki1969}. 
The setting here is a finite bipartite chain $A=[-N/2,a]$, $B=[a+1,N/2]$ with a cut after some site $a$.
\begin{lemma}\label{lem:1D}
Let $H=\sum_{i=-N/2}^{N/2}h_i$ be a 1D Hamiltonian with $h_i$ having support on at most $l$ continuous sites. 
Define $H_A=\sum_{\supp(h_i)\subset A} h_i$ and $H_B$ analogously.
Then,
\beq
\max\{\|E_\beta\|,\|E_\beta^{-1}\|\}\le\exp\left(\frac 1 2 f(\beta, J, l)\exp(f(\beta,J,l))\right),
\enq
where $f(\beta, J, l)=4\beta Jl^2e^{1+4\beta Jl}$.
\end{lemma}
This bound is no longer linear in $\beta$ and the interaction strength, but it still proves the following area law, as it is independent of the size of $A$ and $B$.
\begin{theorem}[One dimension]\label{thm:1DAL}
In 1D, the R\'enyi mutual information is bounded as:
\beq
I_\infty(A:B)\le 4f(\beta,J,l)\exp(f(\beta,J,l))
\enq
\end{theorem}

In higher dimensions, we can use an imaginary-time Lieb-Robinson bound for high temperatures \cite{bratteli2012operator,abanin2015exponentially,arad2016,kuwahara2016}.
\begin{lemma}
\label{lem:highTD}
On a $D$-dimensional lattice, we have
\beq
\log(\|E_\beta\|)\le -\log(1-2\beta J k)|\partial A|/(2k)
\enq
if $2\beta Jk<1$ and the same bound holds for $\|E_\beta^{-1}\|$.
\end{lemma}
We give the proof in Appendix \ref{sec:highTD}. This results in our last area law based on Lemma \ref{lem:geometric}:
\begin{theorem}[Any dimension at high temperature]\label{thm:DDAL}
In the setting of the previous lemma and for $\beta J k<1$, we have
\beq
I_\infty(A:B)\le -4\log(1-\beta J k)|\partial A|/k.
\enq
\end{theorem}

Both Lemmas \ref{lem:1D} and \ref{lem:highTD}, use an imaginary time-ordered integral to bound the norm of $E_\beta$, as
\beq
\begin{split}
&\vert \vert E_\beta \vert \vert =\|e^{-\beta(H_A+H_B)}e^{\beta H}\|\\
&\le\exp\left(\int_0^\beta \|e^{x(H_A+H_B)}H_Ie^{-x(H_A+H_B)}\|\text{d}x\right),
\end{split}
\enq
which follows from the Dyson series and the triangle inequality.
The same bound holds for the inverse operator $E_\beta^{-1}$ with $(H_A+H_B)$ replaced by $H$ on the right-hand side.
The operator in the norm corresponds to an imaginary time evolution of $H_I$, which can be approximated with a Taylor series.

This technique cannot be used to extend the area law to arbitrary dimensions and temperatures because there exists a 2D lattice Hamiltonian such that for sufficiently small temperatures the quantity $\| e^{x H} A e^{-x H} \|/\|A \|$ diverges in the thermodynamic limit \cite{bouch2015,avdoshkin2020euclidean}. As pointed out in \cite{avdoshkin2020euclidean} this bound can be extended to the Bethe lattice, which can be seen as an intermediate case between one and higher dimension. 

{As already mentioned, the previous theorems also extend to all measured R\'enyi mutual informations.}

Let us now comment on the thermal area law for classical systems.
For them, an area law for the mutual information independent of temperature and energy was shown in \cite{wolf2008}. 
The idea behind it is the Markov property of classical thermal states, which reads $P(x_A|x_B)=P(x_A|x_{\partial B})$, where again the boundaries are defined as above such that there is no interaction between $A$ and $B\setminus\partial B$ and vice-versa. 
This means that all correlations between $A$ and $B$ are mediated through $\partial B$ and also implies that the correlations cluster at the boundary, in the sense that $I(A:B)=I(\partial A:\partial B)$. This leads to a bound that only depends on the dimension of the boundary.

This latter equality also holds for the R\'enyi mutual information defined using R\'enyi divergences, which allows for the following extension.
\begin{theorem}[Classical temperature-independent thermal area law]
\label{thm:clAreaLaw}
For a classical system with local dimension $d$, we have
\beq
I_\alpha^\mathbb{M}(A:B)\le (|\partial A|+|\partial B|)\log d
\enq
for $\alpha\in(0,1)\cup(1,2]$.
\end{theorem}
{Note that in the classical case the measured mutual information coincides with the analogous definition from Eq. \eqref{eq:renyiDivClassical}.}
The proof, which uses the fact that every probability distribution majorizes the flat distribution, is given in Appendix \ref{sec:clAreaLaw}.
A challenge in the R\'enyi case is that for fixed system size the mutual information is no longer bounded in general, but only for $\alpha\le2$.
A simple example for two bits shows that it can be arbitrarily large for $\alpha>2$, and by extension for $I_\infty$ (see Appendix \ref{sec:clAreaLaw}), in which case we can only give the temperature-dependent Theorem \ref{thm:commuting}.

\section{R\'enyi mutual information on tensor network states}\label{sec:PEPO}
Matrix product states and also their higher dimensional analog projected entangled pair states (PEPS) \cite{verstraete2004}, have by construction a small bipartite entanglement entropy, bounded by the logarithm of the bond dimension $D$ times the number of neighboring pairs across the boundary of the bipartition. The same holds for their mutual information, as for pure states it is equal to twice their entanglement entropy.

A natural question is whether this extends to the mutual information of projected entangled pair density operators (PEPDOs), their mixed state analog \cite{cirac2020}.
In \cite{wolf2008}, this question was answered positively for the mutual information $I(A:B)$, using the additional assumption that the PEPDO has a \textit{local purification}. This means that there exists a PEPS with a physical and an ancilla index of equal dimension on every site, whose partial trace over the ancillas equals the PEPDO\footnote{Notice that not all PEPDOs admit such description \cite{de2013purifications,de2016fundamental}.}.

This result can be extended to the measured R\'enyi mutual information for a limited range of $\alpha$ (see Appendix \ref{sec:mutPure}):
\begin{theorem}\label{thm:PEPO}
For a PEPDO with local purification, bond dimension $D$, and $|\partial A|$ the number of bonds between $A$ and $B$, it holds
\beq
I^\mathbb{M}_\alpha(A:B)\le2|\partial A|\log D
\enq
for $\alpha\in(0,1)\cup(1,2]$.
\end{theorem}
This can be proven by first noticing that the problem is equivalent to the pure state case if one  considers the purification.
Then, the trace over the ancillas of the PEPS, which yields the PEPDO, does not increase the mutual information as it is a local operation. The remaining step is to compute the mutual information of a pure state $\rho=\ketbran{\psi}\psi$ and relate it to an entropy of the subsystems. This is similar to the von Neumann case, where $I(A:B)=2S(\rho_A)$.
The entropy of the marginal is then bounded by the logarithm of the number of Schmidt values of the decomposition into $A$ and $B$, which yields the desired bound.
We give an analog proof for the R\'enyi mutual information in Appendix \ref{sec:mutPure} valid for $\alpha\le2$.
For $\alpha>2$ we present a simple counterexample on 2 qubits ($D=|\partial A|=2$) with arbitrarily large measured R\'enyi mutual information in the appendix.

\section{Correlation functions}\label{sec:corr}
The mutual information quantifies both classical and quantum correlations \cite{groisman2005}.
Therefore, it seems intuitive that it should also impose a bound on correlation functions, given by $\mathcal C(M_A,M_B):=\langle M_A M_B\rangle-\langle M_A\rangle\langle M_B\rangle$ for observables $M_A$, $M_B$ supported on $A$, $B$ respectively.
This was established for $I(A:B)$ in \cite{wolf2008}. We here extend this proof to the R\'enyi case, by showing that
\beq\label{eq:corr}
I_\alpha^\mathbb{M}(A:B)\ge\frac{\min\{1,\alpha\}\mathcal{C}(M_A,M_B)^2}{2\|M_A\|^2\|M_B\|^2} .
\enq
The bound trivially extends to all other quantum R\'enyi divergences that fulfill the data-processing inequality.

The key technical result is a generalization of the quantum \textit{Pinsker's inequality}:
\begin{lemma}\label{lem:pinsker}
For $\alpha\in (0,1)\cup(1,\infty)$ we have
\beq
\min\{1,\alpha\} \frac 1 2 \|\rho-\sigma\|_1^2 \le D_\alpha^\mathbb{M}(\rho\|\sigma).
\enq
\end{lemma}
The proof in Appendix \ref{sec:pinsker} uses the same argument as for the relative entropy $D(\rho \| \sigma)$, where the data-processing inequality is applied to a binary measurement \cite{hiai1981}.
The known equivalent classical result \cite{gilardoni2010} can then be applied to the measurement outcome.
The bound on the correlations Eq.~\eqref{eq:corr} follows using $\|X\|_1\ge\Tr[XY]/\|Y\|$, exactly as in \cite{wolf2008}.

\section{Conclusion}We have given alternative definitions of the mutual information. We have shown that, as a measure of bipartite correlations, they satisfy a number of desirable properties, including area laws for thermal states.

{As a main advantage over the von Neumann mutual information, we provide DMRG like algorithms to compute these quantities. This should help to characterize correlations in mixed states in a more rigorous way than with the previously used $I_\alpha$. It would be interesting for future numerical studies to evaluate the performance of this algorithm. }

These R\'enyi mutual information measures differ from the widely used definition of Eq. \eqref{eq:ialpha}, which has nonetheless found applications in, for instance, analyzing thermal phase transitions \cite{singh2011finite}. Our results do not rule out the possibility that $I_\infty(A:B)$ has a wildly different behavior (such as a volume law) at thermal criticality, which occurs at low temperatures when $D>1$. It would be interesting to study whether singularities in it appear at phase transition points or an area law holds with full generality.

These quantities may be a useful measure of correlations beyond thermal states, such as in dissipative dynamics, quantum quenches, and non-equilibrium steady states. We hope that our results motivate their study in the wider context of quantum many-body systems. An interesting future question is whether the smallness of any of these quantities guarantees an efficient approximation of mixed states via tensor networks, similar to how the R\'enyi entanglement entropy guarantees MPS approximations in the pure state case \cite{verstraete2006matrix}. The R\'enyi \textit{entanglement of purification} is known to play such a role \cite{Guth_Jarkovsk__2020}, but is hard to calculate in practice.

\begin{acknowledgments}
The authors acknowledge useful discussions with \'Angela Capel and a useful comment by Aleksandr Avdoshkin. \'A.M.A, G.S., and J.I.C. acknowledge funding from ERC Advanced Grant QUENOCOBA under the EU Horizon 2020 program (Grant Agreement No. 742102) and within the D-ACH Lead-Agency Agreement through project No. 414325145 (BEYOND C).
\end{acknowledgments}

\bibliographystyle{apsrev4-2}
\bibliography{references}

\appendix
\widetext

\section{Quantum R\'enyi divergences}\label{sec:divergences}

In this section we recall various definitions and facts for the different R\'enyi divergences, which will be useful for the proofs in the following sections.

The \textit{Petz R\'enyi divergence}~\cite{petz1986} is perhaps the most direct quantum analogue of the classical family of R\'enyi divergences, that is, for density operators that do not, in general, commute. It is defined for $\alpha\in(0,1)\cup(1,2]$ as
\beq
\overline D_\alpha(\rho\|\sigma)=\frac{1}{\alpha-1}\log\Tr[\rho^\alpha\sigma^{1-\alpha}]
\enq
if $\supp(\rho)\subseteq\supp(\sigma)$ and $+ \infty$ otherwise, where the inverse of $\sigma$ is understood to be a pseudo-inverse. 
While this seems to be a straightforward generalization, several other definitions that collapse to the classical definition for the commuting case are possible.

We will invoke two main alternatives for the purposes of this work. They are the \textit{sandwiched R\'enyi divergence}~\cite{muellerlennert2013,wilde2014strong}, which for $\alpha \in [1/2,1)\cup(1,\infty)$ takes the form
\beq
\widetilde D_\alpha(\rho\|\sigma)=\frac{1}{\alpha-1}\log\Tr\left[\left(\sigma^{\frac{1-\alpha}{2\alpha}}\rho\sigma^{\frac{1-\alpha}{2\alpha}}\right)^\alpha\right]
\enq
and the \textit{geometric R\'enyi divergence}~\cite{matsumoto2018}, which for $\alpha \in (0,1)\cup(1,\infty)$ reads
\beq
\widehat D_\alpha(\rho\|\sigma)=\frac{1}{\alpha-1}\log\Tr\left[\sigma\left(\sigma^{-1/2}\rho\sigma^{-1/2}\right)^\alpha\right].
\enq
This geometric R\'enyi divergence is crucial in our proof of the thermal area law.

These quantities share some important properties as being increasing in $\alpha$
\beq
D_\alpha(\rho\|\sigma)\le D_\beta(\rho\|\sigma) \text{ for }\alpha<\beta,
\enq
nonnegativity
\beq
D_\alpha(\rho\|\sigma)\ge0,
\enq
and a data-processing inequality 
\beq
D(\mathcal E(\rho)\|\mathcal E(\sigma))\le D(\rho\|\sigma) \text{ for any CPTP map } \mathcal E.
\enq
While for the sandwiched and Petz R\'enyi divergence this data-processing inequality holds for the full range of $\alpha$ given above, for the geometric, this is only known for $\alpha\le2$.
Additionally, we have the inequalities
\begin{equation}\label{eq:divergenceIneqs}
\begin{aligned}
\widetilde D_\alpha(\rho\|\sigma) \le \overline D_\alpha(\rho\|\sigma) &\text{ for } \alpha\in[1/2,1)\cup(1,2],\\
\overline D_\alpha(\rho\|\sigma) \le \widehat D_\alpha(\rho\|\sigma) &\text{ for } \alpha\in(0,1)\cup(1,2],\\
\widetilde D_\alpha(\rho\|\sigma) \le \widehat D_\alpha(\rho\|\sigma) &\text{ for }\alpha\in[1/2,0)\cup(1,\infty).
\end{aligned}
\end{equation}
The sandwiched and Petz R\'enyi divergences converge to the Umegaki relative entropy in the limit $\alpha \to 1$, while the geometric converges to the so-called \textit{Belavkin-Staszewski relative entropy} \cite{belavkin1982}.
For $\alpha \to\infty$, the sandwiched and geometric R\'enyi divergences converge to the maximal R\'enyi divergence
\beq
D_\infty(\rho\|\sigma)=\log\inf\{\lambda: \rho\le\lambda\sigma\}.
\enq
This is the main reason for using the geometric R\'enyi divergence as a tool in many of the following proofs: the statements for $D_\infty$ are then derived by taking the limit $\alpha\to\infty$.
For a more complete summary of the properties of R\'enyi divergences and their proofs, see~\cite{khatri2020} and the references therein.

In the main text we listed several expressions for the mutual information:
\begin{subequations}
\begin{align}
I(A:B) &=  S(\rho_A) +S(\rho_B) - S(\rho_{AB})\label{eq:mutSapp}\\
&=D(\rho_{AB}\| \rho_A\otimes \rho_B)\label{eq:mutDapp}\\
&=\min_{\sigma_B}D(\rho_{AB}\|\rho_A\otimes \sigma_B)\label{eq:mutOptapp}
\end{align}
\end{subequations}
We can now introduce R\'enyi mutual information arising from Eq.~\eqref{eq:mutDapp}:
\beq
\widehat I_\alpha(A:B)=\widehat D_\alpha(\rho_{AB}\|\rho_A\otimes\rho_B)
\enq
and $\overline I_\alpha$ and $\widetilde I_\alpha$ respectively.
We note that all upper bounds for $\widehat I(A:B)$ are automatically also upper bounds to $\widetilde I$ and $\overline I$ due to the inequalities \eqref{eq:divergenceIneqs}.

In the quantum information literature the mutual information is commonly defined in a different, inequivalent way:
Instead of starting from Eq. \eqref{eq:mutDapp}, Eq. \eqref{eq:mutOptapp} is taken, and the Umegaki relative entropy is being replaced by a R\'enyi divergence:
\beq \label{eq:mutOptRenyi}
\widehat I_\alpha^{opt}(A:B)=\min_{\sigma_B} \widehat D_\alpha(\rho_{AB}\|\rho_A\otimes\sigma_B)
\enq
Again, a variety of definitions can be made by choosing the different R\'enyi divergences introduced above.
There are operational interpretations in information theory for these quantities, for example in quantum hypothesis testing \cite{hayashi2016} or entanglement-assisted single-shot communication protocols \cite{anshu2017}. Since they contain an additional optimization, these quantities are more difficult to compute from a practical standpoint, and we do not use them in this paper. However, all upper bounds (such as the area laws) trivially hold for them as well.

\section{Negative R\'enyi mutual information of a classical Ising chain}
\label{sec:ising}
In this section, we present the calculation of the R\'enyi mutual information based on R\'enyi entropies
\beq
I_\alpha(A:B) = S_\alpha(\rho_A) +S_\alpha(\rho_B) - S_\alpha(\rho_{AB})
\enq
for a classical Ising chain in its thermal state and show that it can become negative.
We adapt the standard method of transfer matrices used to solve the classical Ising spin chain (see, e.g., \cite{friedli2017}).
The model consists of $N$ spins $z_i$ taking values $\pm 1$ and is placed on a ring, i.e., we use periodic boundary conditions.
The Hamiltonian is given by
\beq
H=\sum_{i=1}^N hz_i + Jz_iz_{i+1},
\enq
with the temperature included in the constants and the addition understood as modulo $N$.
A general technical problem occurring in the calculation of mutual information is the breaking of translation symmetry due to the definition of distinct regions $A$ and $B$.
In order to deal with this problem, we put two subsystems $A={1,\cdots,L}$ and $B={L+1,\cdots, 2L}$ on a chain of length $N>2L$ and calculate the mutual information in the limit of first taking $N\to\infty$ and then $L\to\infty$. 
We define the matrix
\beq \label{eq:transferDef1}
T=\left( \begin{array}{rr}
e^{-J+h}&e^J\\
e^J&e^{-J-h}
\end{array}\right)= \lambda_+\ketbran{\lambda_+}{\lambda_+}+\lambda_-\ketbran{\lambda_-}{\lambda_-}
\enq
with $\lambda_+>\lambda_-$ the eigenvalues and $\ket{\lambda_+}$, $\ket{\lambda_-}$ the corresponding eigenvectors.
Note that the Perron-Frobenius theorem guarantees that $\lambda_+$ is the unique largest eigenvalue (in norm) and also positive~\cite{perron1907}.
The computational basis is denoted by $\ket{\pm 1}$.
We first calculate the partition function by rewriting it with a matrix power
\begin{equation}
\begin{aligned}
\mathcal{Z} &= \sum_{\sigma_1,\cdots,\sigma_N\in\{-1,1\}} \prod_{i=1}^N \exp(-h\frac{\sigma_i+\sigma_{i+1}}{2}-J\sigma_i\sigma_{i+1})\\
		&= \sum_{\sigma_1,\cdots,\sigma_N\in\{-1,1\}} \prod_{i=1}^N \bra{\sigma_i}T\ket{\sigma_{i+1}} = \Tr\left[T^N\right] = \lambda_+^N+\lambda_-^N
\end{aligned}
\end{equation}
and use this result to calculate probabilities as follows. 
\begin{equation}
\begin{aligned}
P(z_k=\pm1) &= \mathcal{Z}^{-1}\sum_{\sigma_1,\cdots,\sigma_N\in\{-1,1\}} \delta_{\pm1,\sigma_k} \prod_{i=1}^N \exp(-h\frac{\sigma_i+\sigma_{i+1}}{2}-J\sigma_i\sigma_{i+1})\\
		&=\mathcal{Z}^{-1} \sum_{\sigma_1,\cdots,\sigma_N\in\{-1,1\}} \bra{\sigma_1}T \ket{\pm1}\braketn{\pm1}{\sigma_2}\prod_{i=2}^N \bra{\sigma_i}T\ket{\sigma_{i+1}} \\
		&= \mathcal{Z}^{-1}\Tr\left[T^N\ket{\pm1}\bra{\pm1}\right]\\
		&=\frac{\lambda_+^N \braketn{\lambda_+}{\pm1}\braketn{\pm1}{\lambda_+}+\lambda_-^N\braketn{\lambda_-}{\pm1}\braketn{\pm1}{\lambda_-}}{\lambda_+^N+\lambda_-^N} \xrightarrow{N\rightarrow\infty} \braketn{\lambda_+}{\pm1}\braketn{\pm1}{\lambda_+}=\braketn {\pm1} {\lambda_+}^2 ,
\end{aligned}
\end{equation}
where in the above we used translation invariance and cyclicity of the trace.
We now generalize this calculation to the probability of a configuration of several spins and apply it to a conditional probability.
\begin{equation}
\label{eq:condProb}
\begin{aligned}
P(z_k=\sigma_k&|z_{k-1}=\sigma_{k-1},\cdots,z_1=\sigma_1)\\
&=\frac{\Tr\left[T^{N-k+1}\ketbran{\sigma_k}{\sigma_k}T\ket{\sigma_{k-1}}\cdots\bra{\sigma_2}T\ketbran{\sigma_1}{\sigma_1}\right]}{\Tr\left[T^{N-k+2}\ketbran{\sigma_{k-1}}{\sigma_{k-1}}T\ket{\sigma_{k-2}}\cdots\bra{\sigma_2}T\ketbran{\sigma_1}{\sigma_1}\right]}\\
&=\frac{\Tr\left[T^{N-k+1}\ketbran{\sigma_k}{\sigma_k}T\ketbran{\sigma_{k-1}}{\sigma_1}\right]}{\Tr\left[T^{N-k+2}\ketbran{\sigma_{k-1}}{\sigma_1}\right]}\\
&=\bra{\sigma_k}T\ket{\sigma_{k-1}}\frac{\bra{\sigma_1}T^{N-k+1}\ket{\sigma_k}}{\bra{\sigma_1}T^{N-k+2}\ket{\sigma_{k-1}}}\\
&=\bra{\sigma_k}T\ket{\sigma_{k-1}}\frac{\braketn{\sigma_1}{\lambda_+}\braketn{\lambda_+}{\sigma_k}\lambda_+^{N-k+1}+\braketn{\sigma_1}{\lambda_-}\braketn{\lambda_-}{\sigma_k}\lambda_-^{N-k+1}}{\braketn{\sigma_1}{\lambda_+}\braketn{\lambda_+}{\sigma_{k-1}}\lambda_+^{N-k+2}+\braketn{\sigma_1}{\lambda_-}\braketn{\lambda_-}{\sigma_{k-1}}\lambda_-^{N-k+2}}\\
&\xrightarrow{N\to\infty}\bra{\sigma_k}T\ket{\sigma_{k-1}}\frac{\braketn{\lambda_+}{\sigma_k}}{\braketn{\lambda_+}{\sigma_{k-1}}\lambda_+}=P(z_k=\sigma_k|z_{k-1}=\sigma_{k-1}).
\end{aligned}
\end{equation}
In the second equality, the cancellations within the trace explcitly verify the Markov property of the periodic chain. 
In the limit, the dependency on $\sigma_1$ also vanishes, which can be understood from the perspective of correlations decaying with the distance. In the large $N$ limit the latter have to mediate through an infinitely long region of the chain, and hence vanish.

\begin{figure}[t]
\centering
\begin{subfigure}
	\centering
	\includegraphics[width = 0.4\linewidth]{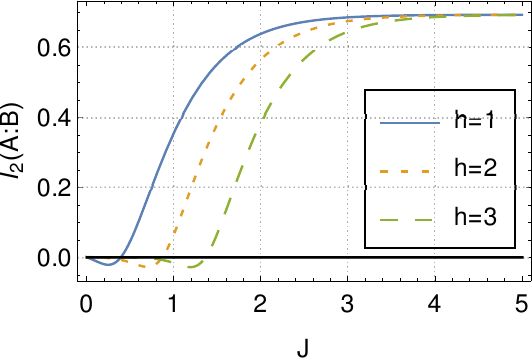}
\end{subfigure}
\begin{subfigure}
	\centering
	\includegraphics[width = 0.4\linewidth]{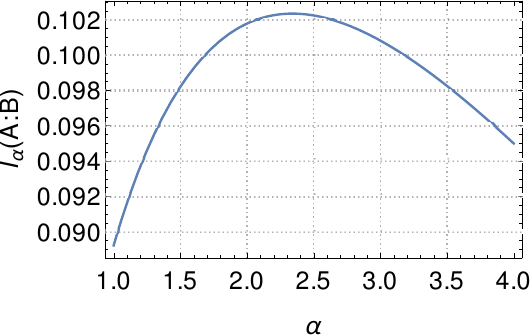}
\end{subfigure}
\caption{Plot of the R\'enyi mutual information $I_\alpha(A:B)$ (Eq.~\eqref{eq:ialpha}) for a classical Ising chain in the limit $L\to\infty$ (Eq. \eqref{eq:isingMutFinal}). (left) 
As a function of $J$ for various values of the field $h$ with $\alpha=2$ fixed. For sufficiently small values of $J$ the mutual information becomes negative. (right) 
As a function of $\alpha$ for $J=0.6$ and $h=1$, demonstrating that the mutual information is not a monotonous function of $\alpha$.}
\label{fig:Imut}
\end{figure}

We are now ready to calculate the R\'enyi entropies
\beq
\label{eq:alphaProbSum}
S_\alpha(A)=\frac{1}{1-\alpha}\log\left(\sum_{\sigma_1,\cdots,\sigma_L\in\{-1,1\}} (P(z_1=\sigma_1)P(z_2=\sigma_2|z_1=\sigma_1)...P(z_L=\sigma_L|z_{L-1}=\sigma_{L-1}))^\alpha\right)
\enq
where we already used \eqref{eq:condProb} to decompose the joint probability. 
Again, due to translation invariance, all the conditional probabilities are the same.
We define
\beq\label{eq:transferDef2}
	T_\alpha=\left(\begin{array}{rr}
P^\alpha(z_k=-1|z_{k-1}=-1)&P^\alpha(z_k=-1|z_{k-1}=1)\\
P^\alpha(z_k=1|z_{k-1}=-1)&P^\alpha(z_k=1|z_{k-1}=1)\end{array}\right)
\textrm{, }\ket{\id}=\left(\begin{array}{r}1\\1\end{array}\right)
\textrm{, }\ket{P_\alpha}=\left(\begin{array}{r}P^\alpha(z_k=-1)\\P^\alpha(z_k=1)\end{array}\right),
\enq
where $T_\alpha$ is no longer a stochastic matrix.
With the same technique as before, \eqref{eq:alphaProbSum} becomes
\beq
S_\alpha(A)=\frac{1}{1-\alpha}\log\left(\bra{\id}T_\alpha^{L-1}\ket{P_\alpha}\right)
\enq
and we obtain the R\'enyi mutual information
\beq
I_\alpha(A:B)=\frac{1}{1-\alpha}\log\left(\frac{\bra{\id}T_\alpha^{L-1}\ket{P_\alpha}^2}{\bra{\id}T_\alpha^{2L-1}\ket{P_\alpha}}\right).
\enq
To evaluate this expression, we use the diagonalization of $T_\alpha$, which reads 
\beq
\left(\begin{array}{rr}
t_\alpha^+&0\\
0&t_\alpha^-\end{array}\right)
=DT_\alpha D^{-1} 
\enq
for some invertible matrix D. 
The fact that $T_\alpha$ is diagonalizable again results from the Perron-Frobenius theorem, which states that the largest eigenvalue has multiplicity one if the matrix has positive entries.
Therefore, another eigenvalue exists, which proves diagonalizability.
Additionally $t_\alpha^+$, which we define to be the larger real eigenvalue, is also strictly larger in absolute value, which allows us to calculate the limit $L\to\infty$ of the above expression.
\begin{equation} \label{eq:isingMutFinal}
\begin{aligned}
	I_\alpha(A:B)&=\frac{1}{1-\alpha}\log\frac{(\bra{\id}D^{-1}\ket{-1}\bra{-1}D\ket{P_\alpha} (t_\alpha^+)^{L-1}+\bra{\id}D^{-1}\ket{1}\bra{1}D\ket{P_\alpha} (t_\alpha^-)^{L-1})^2}{\bra{\id}D^{-1}\ket{-1}\bra{-1}D\ket{P_\alpha} (t_\alpha^+)^{2L-1}+\bra{\id}D^{-1}\ket{1}\bra{1}D\ket{P_\alpha} (t_\alpha^-)^{2L-1}}\\
	&\xrightarrow{L\to\infty} \frac{1}{1-\alpha}\log \frac{\bra{\id}D^{-1}\ket{-1}\bra{-1}D\ket{P_\alpha}}{t_\alpha^+}
\end{aligned}
\end{equation}
Using the eigenvalues and eigenvectors of $T_\alpha$ (Eq. \eqref{eq:transferDef2}) and $T$ (Eq. \eqref{eq:transferDef1}) one can easily derive an explicit analytic expression for the R\'enyi mutual information $I_{\alpha}(A:B)$. However, we omit writing the exact expression for simplicity, due to its length.
Instead we numerically evaluate the formula with the resulting plots in Figure~\ref{fig:Imut}, which show the existence of a negative regime for antiferromagnetic coupling.
Additionally, we find that the mutual information is not monotonous in $\alpha$.
As mentioned in the main text, the existence of a negative regime also proves that the mutual information violates the nonincrease under local operations, because tracing out the $A$ system is a local quantum operation that increases $I_\alpha(A:B)$ to zero.

\section{Technical proofs of main results}

\subsection{Optimizer for the measured R\'enyi-divergence with \texorpdfstring{$\alpha=2$}{a}}

\label{sec:optimizer}
For $\alpha=2$ the expression from the main text for the measured R\'enyi divergence becomes
\beq
D^\mathbb{M}_2(\rho\|\sigma)=\log\sup_{\omega>0} 2\Tr\left[\rho\omega\right]-\Tr\left[\sigma\omega^2\right]
\enq
with $\sigma$ a state with full support.
From \cite{berta2017}, it is known that this expression has an optimizer. 
Here, we prove the explicit expression
\beq
\ket{\omega}=\frac{1}{2} \frac{1}{\sigma\otimes\id+\id\otimes\sigma}\ket{\rho}
\enq
for the optimizer.
We use a vectorized notation with the mapping $\ketbra ij \mapsto \ketn i\ketn j$
We denote the target function by $f(\omega)=2\Tr\left[\rho\omega\right]-\Tr\left[\sigma\omega^2\right]$.
We can extend the supremum over all hermitian operators, which does not change its value, because for any positive semidefinite operators $\omega^+$, $\omega^-$ we have $f(\omega^+-\omega^-)\le f(\omega^+)$.

Given the optimizer $\omega$, the linear term of $f(\omega+\delta\omega)$ in $\delta\omega$ must vanish as $f$ is differentiable.
\beq
\begin{aligned}
f(\omega+\delta\omega)-f(\omega)&=2\Tr[\rho\delta\omega]-\Tr[\sigma\omega\delta\omega+\sigma\delta\omega\omega]+\mathcal{O}(\delta\omega^2)\\
&=\Tr[(2\rho-\sigma\omega-\omega\sigma)\delta\omega]+\mathcal O (\delta\omega^2)
\end{aligned}
\enq
This vanishes for any $\delta\omega$ if and only if $2\rho-\sigma\omega-\omega\sigma=0$. The solution to this linear equation can either be written using the invertible superoperator $\Phi(\omega)=\sigma\omega+\omega\sigma$ as
\beq
\omega=\frac12\Phi^{-1}(\rho)
\enq
or in vectorized notation as in Eq. \eqref{eq:optimizer}
where also the invertibility of $\Phi$ for a $\sigma$ with full rank becomes apparent.

\subsection{Proof of Lemma \ref{lem:geometric}}\label{sec:geometric}
{We restate the Lemma from the main text:

\textbf{Lemma 1. }\textit{Let $E_\beta=e^{-\beta (H_A+H_B)}e^{\beta H}$. For any Hamiltonian defined as in Section \ref{sec:areaLaw}, the maximal R\'enyi mutual information of a thermal state fulfills
\beq
I_\infty(A:B)\le\beta \|H_I\|+\log\left(\|E_{\beta/2}\|^2 \|E_{\beta/2}^{-1}\|^4\right).
\enq}}
The strategy we will follow in order to achieve an upper bound on $D_\infty$, is to first prove the following $\alpha$-independent bound for the geometric divergence, 
\beq
\label{eq:geomFinal}
\widehat D_\alpha(\rho_{AB}\|\rho_A\otimes\rho_B)\le\log\left(\|E_{\beta/2}\|^2\|E_{\beta/2}^{-1}\|^4\frac{Z}{Z_AZ_B}\right)
\enq
with $\rho=\exp(-\beta H)/Z$, $\rho_A=\Tr_B[\rho]$ and $\rho_B=\Tr_A[\rho]$. Then one can take the trivial limit $\alpha \to \infty$ and deduce that the same bound holds for $D_{\infty}$

We start with the argument of the logarithm in $\widehat D$ and assume $\alpha>1$ to be an integer:
\begin{equation}
\begin{aligned}
\label{eq:geomTrace1}
\Tr\left[(\rho_A\otimes\rho_B)\left((\rho_A\otimes\rho_B)^{-1/2}\rho_{AB}(\rho_A\otimes\rho_B)^{-1/2}\right)^\alpha\right] &=\Tr\left[(\rho_{AB})\left((\rho_{AB})^{1/2}(\rho_A\otimes\rho_B)^{-1}(\rho_{AB})^{1/2}\right)^{\alpha-1}\right]\\
&\le\left\|(\rho_{AB})^{1/2}(\rho_A\otimes\rho_B)^{-1}(\rho_{AB})^{1/2}\right\|^{\alpha-1}\\
\end{aligned}
\end{equation}
We used cyclicity of the trace, submultiplicativity of the operator norm, and the inequality
\beq
\label{eq:TrHold}
\Tr[AB]=\Tr[B^{1/2}AB^{1/2}]=\|B^{1/2}AB^{1/2}\|_1\le\|B^{1/2}\|^2\|A\|_1=\|B\|\Tr[A]
\enq
using Hölder's inequality for positive operators $A$ and $B$.
From \eqref{eq:geomTrace1}, we continue with
\begin{equation}
\label{eq:geomTrace2}
\begin{aligned}
\Tr&\left[(\rho_A\otimes\rho_B)((\rho_A\otimes\rho_B)^{-1/2}\rho_{AB}(\rho_A\otimes\rho_B)^{-1/2})^\alpha\right]\\
&\le\|e^{-\beta H/2}e^{\beta(H_A+H_B)/2}\|^{2(\alpha-1)} \|e^{-\beta(H_A+H_B)/2}(\rho_A\otimes\rho_B)^{-1}\frac{1}{Z_AZ_B}e^{-\beta(H_A+H_B)/2}\|^{\alpha-1}\left(\frac{Z_AZ_B}{Z}\right)^{\alpha-1}\\
&=\|e^{-\beta H/2}e^{\beta(H_A+H_B)/2}\|^{2(\alpha-1)} \|(e^{\beta H_A/2}\rho_Ae^{\beta H_A/2}Z_A)^{-1}\|^{\alpha-1}\|(e^{\beta H_B/2}\rho_Be^{\beta H_B/2}Z_B)^{-1}\|^{\alpha-1}\left(\frac{Z_AZ_B}{Z}\right)^{\alpha-1}.
\end{aligned}
\end{equation}
The first norm is already of the desired form. 
For the second and third norms, we write
\beq
\label{eq:subregionNorm}
\begin{aligned}
\|(e^{\beta H_A/2}\rho_Ae^{\beta H_A/2}Z_A)^{-1}\|&=\|\Tr_B[e^{\beta H_A/2}\rho e^{\beta H_A/2}Z_A]^{-1}\|\\
&=\|\Tr_B[e^{\beta(H_A+H_B)/2} e^{-\beta H_{AB}} e^{\beta(H_A+H_B)/2} \frac{e^{-\beta H_B}}{Z_B}\frac{Z_AZ_B}{Z}]^{-1}\|\\
&=\|\Tr_B[e^{\beta (H_A+H_B)/2} e^{-\beta H_{AB}} e^{\beta (H_A+H_B)/2} \frac{e^{-\beta H_B}}{Z_B}]^{-1}\| \frac{Z}{Z_AZ_B},
\end{aligned}
\enq
where we used the cyclicity of the partial trace on $B$ with respect to operators supported only in $B$. 
The operator norm is just the inverse of the smallest eigenvalue of the partial trace. 
Let $\ket{\psi_A}$ be the corresponding eigenvector on $A$ to this eigenvalue and $p_i$, $\ketn{\phi_i}$ an eigensystem of $\exp(-\beta H_B)/Z_B$. Then we get
\begin{equation}
\begin{aligned}
\bra{\psi_A}&\Tr_B[e^{\beta (H_A+H_B)/2}e^{-\beta H_{AB}}e^{(\beta H_A+H_B)/2}\rho_B]\ket{\psi_A}\\
&=\bra{\psi_A}\sum_i p_i\bra{\phi_i}e^{\beta (H_A+H_B)/2}e^{-\beta H_{AB}}e^{\beta (H_A+H_B)/2}\ket{\phi_i}\ket{\psi_A}\\
&=\sum_i p_i\bra{\psi_A}\bra{\phi_i}e^{\beta (H_A+H_B)/2}e^{-\beta H_{AB}}e^{\beta (H_A+H_B)/2}\ket{\phi_i}\ket{\psi_A}\\
&\ge\sum_i p_i\|e^{-\beta (H_A+H_B)/2}e^{\beta H_{AB}}e^{-\beta (H_A+H_B)/2}\|^{-1}\\
&=\|e^{-\beta (H_A+H_B)/2}e^{\beta H_{AB}}e^{-\beta (H_A+H_B)/2}\|^{-1}
\end{aligned}
\end{equation}
by bounding an expectation value of a positive operator by its minimum eigenvalue. Inserting this into \eqref{eq:subregionNorm} we get
\beq
\label{eq:subRegionNorm2}
\|(e^{\beta H_A/2}\rho_Ae^{\beta H_A/2}Z_A)^{-1}\|\le\|e^{-\beta (H_A+H_B)/2}e^{\beta H_{AB}}e^{-\beta (H_A+H_B)/2}\|^{-1}\frac Z{Z_AZ_B}
\enq
and combining \eqref{eq:geomTrace2} with \eqref{eq:subRegionNorm2} yields
\begin{equation}
\begin{aligned}
\Tr[(\rho_A\otimes\rho_B)((\rho_A\otimes\rho_B)^{-1/2}\rho_{AB}(\rho_A\otimes\rho_B)^{-1/2})^\alpha]\le
\|E_{\beta/2}\|^{2(\alpha-1)}\|(E_{\beta/2})^{-1}\|^{4(\alpha-1)} \left(\frac{Z}{Z_AZ_B}\right)^{\alpha-1},
\end{aligned}
\end{equation}
which completes the proof of \eqref{eq:geomFinal} for any integer $\alpha>1$. 
The previous result extends to any $\alpha>1$ by rounding up to the next integer because of the monotonicity of $\widehat D_\alpha$ in $\alpha$.
The bound \eqref{eq:geomFinal} also holds for $\overline D_\alpha$ in the range $\alpha\in(0,1)\cup(1,2]$ and for $\widetilde D_\alpha$ and $\alpha\in(0,1)\cup(1,\infty)$ just by the inequalities between these R\'enyi divergences.

Finally, to bound the ratio of partition functions $Z/Z_AZ_B$, we repeat a simple proof from \cite{lenci2005}, Lemma 3.6:
\beq\label{eq:ratioPart}
\begin{aligned}
\left|\log\Tr\left[e^{-\beta(H_A+H_B)}\right]-\log\Tr\left[e^{-\beta(H_A+H_B+H_I)}\right]\right|&=\left|\int_0^1\frac d {dt}\log\Tr\left[e^{-\beta(H_A+H_B)- t \beta H_I}\right]dt\right|\\
&\le\int_0^1\left|\frac{\Tr\left[-\beta H_I e^{-\beta(H_A+H_B+t H_I)}\right]}{\Tr\left[e^{-\beta(H_A+H_B+t H_I)}\right]}\right|dt\le\beta \|H_I\|
\end{aligned}
\enq

Together, \eqref{eq:geomFinal} and \eqref{eq:ratioPart} give the desired bound which also directly proves the bound for $D_\infty$ as the right-hand side does not depend on $\alpha$ and the limit $\alpha \rightarrow \infty$ can trivially be taken.

\subsection{Proof of Lemma \ref{lem:highTD}}\label{sec:highTD}
We restate the lemma from the main text. It features a thermal Lieb-Robinson bound originally due to Ruelle \cite{ruelle1999statistical}, which has previously appeared in \cite{bratteli2012operator,abanin2015exponentially,arad2016,kuwahara2016}:\\ \\
\textbf{Lemma 3.} \textit{
On a $D$-dimensional lattice, we have
\beq
\log(\|E_\beta\|)\le -\log(1-2\beta J k)|\partial A|/(2k)
\enq
if $2\beta Jk<1$ and the same bound holds for $\|E_\beta^{-1}\|$.}
\\ \\
As explained in the main text, we start from the expression 
\beq
\label{eq:imdyson}
\vert \vert E_\beta \vert \vert =\|e^{-\beta(H_A+H_B)}e^{\beta H}\|
\le\exp\left(\int_0^\beta \|e^{x(H_A+H_B)}H_Ie^{-x(H_A+H_B)}\|\text{d}x\right)
\enq
and get, using the Baker-Campbell-Hausdorff formula,
\beq
\label{eq:compImEvoBound}
\|e^{x(H_A+H_B)}H_Ie^{-x(H_A+H_B)}\|\le\sum_{m=0}^\infty \frac{x^m}{m!}\text{ad}_{H_A+H_B}^m(H_I)
\enq
with the adjoint action $\text{ad}_Y(X)=[Y,X]$, whose powers yield nested commutators. We bound them by Lemma~3 from Ref.~\cite{kuwahara2016}
\beq
\label{eq:adBound}
\|\text{ad}_{H_A+H_B}^m(H_I)\|\le\sum_{i\in C}\|\text{ad}_{H_A+H_B}^m(h_i)\|\le\sum_{i\in C} \prod_{j=1}^m 2Jkj\|h_i\|\le J|\partial A| (2Jk)^mm!
\enq
with $C=\{i:\supp(h_i)\cap A\ne\emptyset, \supp(h_i)\cap B\ne\emptyset\}$. 
The inequality $\sum_{i\in C} \|h_i\|\le J|\partial A|$ follows from the definitions.
We insert \eqref{eq:adBound} into \eqref{eq:compImEvoBound} and evaluate the geometric series
\beq
\|e^{x(H_A+H_B)}H_Ie^{-x(H_A+H_B)}\|\le\sum_{m=0}^\infty (2xJk)^m J|\partial A| =\frac{J|\partial A|}{1-2xJk},
\enq
which holds if $2xJk<1$. Finally, inserting this into the integral \eqref{eq:imdyson} we obtain
\beq
\log\|E_\beta\| \le\int_0^\beta \frac{J|\partial A|}{1-2xJk}dx=-\log(1-2\beta Jk)|\partial A|/(2k)
\enq
if $2\beta Jk<1$, which finishes the proof. If we replace $E_\beta$ with $E_\beta^{-1}$, the proof can be repeated replacing $H_A+H_B$ with $H=H_A+H_B+H_I$ and $H_I$ with $-H_I$.

\subsection{Proof of Theorem \ref{thm:clAreaLaw}}\label{sec:clAreaLaw}
{\textbf{Theorem 4 }\textit{For a classical system with local dimension $d$, we have
\beq
I_\alpha^\mathbb{M}(A:B)\le (|\partial A|+|\partial B|)\log d
\enq
for $\alpha\in(0,1)\cup(1,2]$.} }

In the classical case, all R\'enyi divergences including the measured one coincide, so we have to show:
\beq
D_\alpha(\rho_{AB}\|\rho_A\otimes\rho_B)\le (|\partial A|+|\partial B|)\log d
\enq
The proof is split into two steps. First, we show that the mutual information between $A$ and $B$ equals the mutual information between the boundaries and second, we give a dimension-dependent bound for the mutual information.
We denote by $X_A$ the random variable of configurations on the system $A$ and use $A^\circ=A\setminus\partial A$.
We prove the first identity
\begin{equation}
\begin{aligned}
D_\alpha(\rho_{AB}\|\rho_A\otimes\rho_B)&=\frac{1}{\alpha-1}\log \sum_{x_{A^\circ},x_{\partial A},x_{B^\circ},x_{\partial B}} P(X_{A^\circ}=x_{A^\circ},X_{\partial A}=x_{\partial A},X_{B^\circ}=x_{B^\circ},X_{\partial B}=x_{\partial B})\\
&\quad\left(\frac{P(X_{A^\circ}=x_{A^\circ},X_{\partial A}=x_{\partial A},X_{B^\circ}=x_{B^\circ},X_{\partial B}=x_{\partial B})}{P(X_{A^\circ}=x_{A^\circ},X_{\partial A}=x_{\partial A})P(X_{B^\circ}=X_{B^\circ},X_{\partial B}=x_{\partial B})}\right)^{\alpha-1}\\
&=\frac{1}{\alpha-1}\log \sum_{x_{A^\circ},x_{\partial A},x_{B^\circ},x_{\partial B}} P(X_{A^\circ}=x_{A^\circ},X_{\partial A}=x_{\partial A},X_{B^\circ}=x_{B^\circ},X_{\partial B}=x_{\partial B})\\
&\quad\text{  }\left(\frac{P(X_{A^\circ}=x_{A^\circ},X_{\partial A}=x_{\partial A}|X_{B^\circ}=x_{B^\circ},X_{\partial B}=x_{\partial B})}{P(X_{A^\circ}=x_{A^\circ},X_{\partial A}=x_{\partial A})}\right)^{\alpha-1}\\
&=\frac{1}{\alpha-1}\log \sum_{x_{A^\circ},x_{\partial A},x_{\partial B}} P(X_{A^\circ}=x_{A^\circ},X_{\partial A}=x_{\partial A},X_{\partial B}=x_{\partial B})\\
&\quad\text{  }\left(\frac{P(X_{A^\circ}=x_{A^\circ},X_{\partial A}=x_{\partial A}|X_{\partial B}=x_{\partial B})}{P(X_{A^\circ}=x_{A^\circ},X_{\partial A}=x_{\partial A})}\right)^{\alpha-1}\\
&=\frac{1}{\alpha-1}\log \sum_{x_{A^\circ},x_{\partial A},x_{\partial B}} P(X_{A^\circ}=x_{A^\circ},X_{\partial A}=x_{\partial A},X_{\partial B}=x_{\partial B})\\
&\quad\text{  }\left(\frac{P(X_{A^\circ}=x_{A^\circ},X_{\partial A}=x_{\partial A},X_{\partial B}=x_{\partial B})}{P(X_{A^\circ}=x_{A^\circ},X_{\partial A}=x_{\partial A})P(X_{\partial B}=x_{\partial B})}\right)^{\alpha-1}\\
&=\frac{1}{\alpha-1}\log \sum_{x_{\partial A},x_{\partial B}} P(X_{\partial A}=x_{\partial A},X_{\partial B}=x_{\partial B})\left(\frac{P(X_{\partial A}=x_{\partial A},X_{\partial B}=x_{\partial B})}{P(X_{\partial A}=x_{\partial A})P(X_{\partial B}=x_{\partial B})}\right)^{\alpha-1}\\
&=D_\alpha(\rho_{\partial A\partial B}\|\rho_{\partial A}\otimes\rho_{\partial B}),
\end{aligned}
\end{equation}
where the second to last line comes from repeating all previous steps on the $A$ side.

For the next step we introduce the short-hand notation $p(x_A,x_B)=P(X_{\partial A}=x_A,X_{\partial B}=x_B)$ and use the dimension $D_{\partial A\partial B}$ of the system supported over $\partial A\cup\partial B$ to find the following bound:
\beq\label{eq:clDimBound}
\begin{aligned}
D_\alpha(\rho_{\partial A\partial B}\|\rho_{\partial A}\otimes\rho_{\partial B})&=\frac{1}{\alpha-1}\log \sum_{x_A,x_B} p(x_A,x_B)\left(\frac{p(x_A,x_B)}{p(x_A)p(x_B)}\right)^{\alpha-1}\\
&\le\frac{1}{\alpha-1}\log \sum_{x_A,x_B} p(x_A,x_B)\left(\frac{p(x_A,x_B)}{p(x_A,x_B)p(x_A,x_B)}\right)^{\alpha-1}\\
&=\frac{1}{\alpha-1}\log \sum_{x_A,x_B} p(x_A,x_B)^{2-\alpha}\\
&=S_{2-\alpha}(\rho_{\partial A\partial B})\\
&\le\log D_{\partial A \partial B} = (|\partial A|+|\partial B|)\log d
\end{aligned}
\enq
The last inequality uses the Schur concavity of R\'enyi entropies and holds for $\alpha\in(0,1)\cup(1,2]$.

At this point, it might be natural to wonder if the above area law for classical thermal states can be extended to the case of $\alpha > 2$. The following counterexample rules out a temperature-independent area law in this range. 
We take two bits $x_A,x_B\in\{0,1\}$ and choose the probability distribution $\diag(\epsilon, 0, 0, 1-\epsilon)$ for a constant $\epsilon$ written as a diagonal density matrix.
The marginal probabilities are then $p(0)=\epsilon$, $p(1)=1-\epsilon$ for both $A$ and $B$.
We find that the term for $x_A=x_B=0$ in the sum in the first line of \eqref{eq:clDimBound} reads $\epsilon^{(2-\alpha)}$ and becomes arbitrarily large for sufficiently small $\epsilon$.

Strictly speaking, this example is no thermal state as it contains zero probabilities, but one can choose a sequence of thermal states for every fixed $\epsilon$ that converges to our example.

\subsection{R\'enyi mutual information for pure states and proof of Theorem \ref{thm:PEPO} }\label{sec:mutPure}
Before proving Theorem \ref{thm:PEPO} we give a technical lemma that might be of interest on its own right.
{\begin{lemma}\label{lem:MIS}
For a pure quantum state $\rho_{AB}=\ketbran\psi\psi$ we have the relations
\beq \label{eq:boundPEPOSW}
\widetilde D_\alpha(\rho_{AB}\| \rho_A\otimes\rho_B)=2S_{2/\alpha-1}(\rho_A)
\enq
for $\alpha\in(0,1)\cup(1,2)$, and
\beq \label{eq:boundPEPOPE}
\overline D_\alpha(\rho_{AB}\| \rho_A\otimes\rho_B)=2S_{3-2\alpha}(\rho_A)
\enq
for $\alpha\in(0,1)\cup(1,3/2)$, i.e.,  the R\'enyi mutual information is given by twice the R\'enyi entanglement entropy for appropriate indices $\alpha$.
\end{lemma}}
The following proof resembles a proof from \cite{khatri2020}, where analogous relations are given for $I_\alpha^{opt}(A:B)$ (see \eqref{eq:mutOptRenyi}), which even show that Theorem \ref{thm:PEPO} can be extended to $I_\infty^{opt}(A:B)$ as they hold for all $\alpha$.

We start with the sandwiched R\'enyi divergence and use the Schmidt decomposition
\beq
\ketn{\psi}=\sum_i\sqrt{\lambda_i}\ketn{e_i}\ketn{f_i},
\enq
which also yields a representation of the reduced density matrix
\beq
\rho_A=\sum_i\lambda_i \ketbran{e_i}{e_i}.
\enq
Additionally, we define
\begin{equation}
\begin{aligned}
\ketn{\zeta}&=\left(\rho_A^{(1-\alpha)/2\alpha}\otimes\rho_B^{(1-\alpha)/2\alpha}\right)\ketn{\psi}\\
&=\sum_{i,j,k} \lambda_i^{(1-\alpha)/2\alpha} \ketbran{e_i}{e_i} \otimes \lambda_j^{(1-\alpha)/2\alpha} \ketbran{f_j}{f_j} \lambda_k^{1/2} \ketn{e_k}\ketn{f_k}\\
&=\sum_i \lambda_i^{\frac{2-\alpha}{2\alpha}}\ketn{e_i}\ketn{f_i},
\end{aligned}
\end{equation}
which we can use to rewrite
\begin{equation}
\begin{aligned}
\widetilde D_\alpha(\rho_{AB}\|\rho_A\otimes\rho_B)&=\frac{1}{\alpha-1}\log\Tr\left[\left( \left(\rho_A^{(1-\alpha)/2\alpha}\otimes\rho_B^{(1-\alpha)/2\alpha}\right)\rho_{AB}\left(\rho_A^{(1-\alpha)/2\alpha}\otimes\rho_B^{(1-\alpha)/2\alpha}\right)\right)^\alpha\right]\\
&=\frac{1}{\alpha-1}\log\Tr\left[\ketn{\zeta}\bran{\zeta}^\alpha\right]\\
&=\frac{1}{\alpha-1}\log\Tr\left[\left(\frac{\ketn{\zeta}\bran{\zeta}}{\braketn{\zeta}{\zeta}}\right)^\alpha\right]+\frac{\alpha}{\alpha-1}\log\braketn{\zeta}{\zeta}\\
&=\frac{\alpha}{\alpha-1}\log\braketn{\zeta}{\zeta}\\
&=\frac{\alpha}{\alpha-1}\log\sum\lambda_i^{\frac{2-\alpha}{\alpha}}\\
&=2S_{2/\alpha-1}(\rho_A).
\end{aligned}
\end{equation}
This concludes the proof of \eqref{eq:boundPEPOSW}.

We prove the relation for the Petz-R\'enyi divergence with a similar calculation using the same Schmidt decomposition of $\ketn{\psi}$ and
\begin{equation}
\begin{aligned}
\ketn{\eta}&=\left(\rho_A^{(1-\alpha)/2}\otimes\rho_B^{(1-\alpha)/2}\right)\ketn{\psi}\\
&=\sum_{i,j,k} \lambda_i^{(1-\alpha)/2} \ketbran{e_i}{e_i} \otimes \lambda_j^{(1-\alpha)/2} \ketbran{f_j}{f_j} \lambda_k^{1/2} \ketn{e_k}\ketn{f_k}\\
&=\sum_i \lambda_i^{\frac{3}{2}-\alpha}\ketn{e_i}\ketn{f_i},
\end{aligned}
\end{equation}
and obtain
\begin{equation}
\begin{aligned}
\overline D_\alpha(\rho_{AB}\|\rho_A\otimes\rho_B)&={\frac{1}{\alpha-1}\log\Tr\left[\left(\rho_A^{1-\alpha}\otimes\rho_B^{1-\alpha}\right)\rho_{AB}^\alpha\right]}\\
&=\frac{1}{\alpha-1}\log\Tr\left[\left(\rho_A^{(1-\alpha)/2}\otimes\rho_B^{(1-\alpha)/2}\right)\rho_{AB}^\alpha\left(\rho_A^{(1-\alpha)/2}\otimes\rho_B^{(1-\alpha)/2}\right)\right]\\
&=\frac{1}{\alpha-1}\log\Tr\left[\ketn{\eta}\bran{\eta}\right]\\
&=\frac{1}{\alpha-1}\log\braketn{\eta}{\eta}\\
&=\frac{1}{\alpha-1}\log\sum\lambda_i^{3-2\alpha}\\
&=2S_{3-2\alpha}(\rho_A).
\end{aligned}
\end{equation}
in the range $\alpha \in (0,1)\cup(1,3/2)$.
This closes the proof of \eqref{eq:boundPEPOPE} and thereby of theorem 5.

We now give an extended version of Theorem \ref{thm:PEPO} from the main text as we obtain the additional statements directly from our proof:
\begin{theorem}\label{thm:PEPOext}
For a PEPDO with local purification, bond dimension $D$, and $|\partial A|$ the number of bonds between $A$ and $B$, it holds
\beq\label{eq:PEPOSand}
\widetilde I_\alpha(A:B)\le2|\partial A|\log D
\enq
for $\alpha\in[1/2,1)\cup(1,2]$,
\beq\label{eq:PEPOPetz}
\overline I_\alpha(A:B)\le2|\partial A|\log D
\enq
for $\alpha\in(0,1)\cup(1,3/2]$ and
\beq \label{eq:PEPOMeas}
I^\mathbb{M}_\alpha(A:B)\le2|\partial A|\log D
\enq
for $\alpha\in(0,1)\cup(1,2]$.
\end{theorem}
As already explained in the main text, we can, by the local purification assumption, restrict to PEPS due to the nonincrease of the mutual information under local operations. Using Lemma \ref{lem:MIS} and the fact that all R\'enyi entropies are bounded by the logarithm of the dimension, this gives \eqref{eq:PEPOSand}.
The case of the measured R\'enyi mutual information then follows from the data-processing inequality, which holds for $\alpha\ge1/2$.

For the regime of $\alpha<1/2$, we use the same argument together with the Petz-R\'enyi divergence, which fulfills the data-processing inequality for $\alpha\in(0,1)\cup(1,2]$ and we invoke again Lemma \ref{lem:MIS}.

To show that the range of $\alpha$ cannot be extended we give a simple counterexample of two qubits, which have bond dimension $D=2$ and local dimension $d=2$.
The state $\ketn\psi=\sqrt\epsilon\ket0\ket0+\sqrt{1-\epsilon}\ketn1\ketn1$ with a measurement in the computational basis gives the probability distributions $\diag(\epsilon,0,0,1-\epsilon)$ and $\diag(\epsilon^2,\epsilon-\epsilon^2,\epsilon-\epsilon^2,(1-\epsilon)^2)$ for the full system and the product of marginals respectively.
We already showed the divergence of the mutual information of this distribution in section \ref{sec:clAreaLaw}.

\subsection{Proof of Lemma \ref{lem:pinsker}}\label{sec:pinsker}
We prove a generalization of Pinsker's inequality by adapting a proof for the relative entropy from \cite{hiai1981}.
We use the measured R\'enyi divergence, which is not greater than any R\'enyi divergence in the range where the non-measured fulfills the data-processing inequality.
For two quantum states $\rho$ and $\sigma$, define $\phi^+$ to be the projector on the non-negative subspace of $\rho-\sigma$.
Then, we have
\beq
\|\rho-\sigma\|_1= \Tr[(\rho-\sigma)\phi^+]-\Tr[(\rho-\sigma)(\id-\phi^+)]=2(\Tr[\rho\phi^+]-\Tr[\sigma\phi^+]).
\enq
On the other hand, the supremum of the measured R\'enyi divergence can be lower bounded by a particular measurement, which we choose to be $\phi^+, \id-\phi^+$:
\beq
D_\alpha^\mathbb M(\rho\|\sigma)=\sup_{(\chi,M)}(P_{\rho,M}\|P_{\sigma,M})\ge D_\alpha\big((\Tr[\rho\phi^+],\Tr[\rho(\id-\phi^+)])\|(\Tr[\sigma\phi^+],\Tr[\sigma(\id-\phi^+)])\big),
\enq
where the latter expression is to be understood as the R\'enyi divergence of the binary probability distribution.
Using the total variation distance of this binary distribution $V=2\Tr[(\rho-\sigma)\phi^+]$ we can apply the classical Pinsker inequality $D_\alpha\ge \alpha V^2/2$ for $\alpha\in(0,1)$ from \cite{gilardoni2010} to obtain
\beq
D_\alpha^\mathbb M(\rho\|\sigma)\ge\frac{\alpha}2 V^2=\frac{\alpha}2\|\rho-\sigma\|_1^2
\enq
and
\beq
D_\alpha^\mathbb M(\rho\|\sigma)\ge\frac{1}2\|\rho-\sigma\|_1^2
\enq
for $\alpha>1$ by the monotonicity of $D_\alpha^\mathbb{M}$.

\end{document}